\documentclass[
aps,%
11pt,%
final,%
notitlepage,%
oneside,%
twocolumn,%
nobibnotes,%
nofootinbib,%
superscriptaddress,%
noshowpacs,%
centertags]%
{revtex4}

\begin{document}




\title{The EMCCD-Based Speckle Interferometer of the BTA 6-m
Telescope: Description and First Results}

\author{\firstname{A.~F.}~\surname{Maksimov}}
\affiliation{\saoname}

\author{\firstname{Yu.~Yu.}~\surname{Balega}}
\affiliation{\saoname}

\author{\firstname{V.~V.}~\surname{Dyachenko}}
\affiliation{\saoname}

\author{\firstname{E.~V.}~\surname{Malogolovets}}
\affiliation{\saoname}

\author{\firstname{D.~A.}~\surname{Rastegaev}}
\affiliation{\saoname}

\author{\firstname{E.~A.}~\surname{Semernikov}}
\affiliation{Research Institute of Multiprocessor Computing
Systems, Southern Federal University, \\ Taganrog, 347900 Russia}

\received{January 30, 2009}%
\revised{July 14, 2009}%

\begin{abstract}The description is given for the speckle interferometer of the BTA 6-m
telescope of the SAO RAS based on a new detector with an electron
multiplication CCD. The main components of the instrument are
microscope objectives, interference filters and atmospheric
dispersion correction prisms. The PhotonMAX-512B CCD camera using
a back-illuminated CCD97 allows up to 20 speckle images (with
512$\times$512 pix resolution) per second storage on the hard
drive. Due to high quantum efficiency (93\% in the maximum at 550
nm), and high transmission of its optical elements, the new camera
can be used for diffraction-limited (0.02$''$) image
reconstruction of $15^{m}$ stars under good seeing conditions. The
main advantages of the new system over the previous generation BTA
speckle interferometer are examined.

\end{abstract}
\maketitle

\section{INTRODUCTION}

Improvement of the angular resolution of ground-based optical
telescopes remains one of the most important astronomical
problems. The advantages of the largest astronomical instruments
cannot be implemented without an introduction of the newest
methods of correction of phase degradations, arising during the
light propagation through the turbulent atmosphere. This problem
obtains a special meaning in connection with the development of
the new generation of giant telescopes.

The adaptive optics and speckle interferometry allow approaching
the diffraction limit of the telescope's angular resolution. Image
reconstruction in speckle interferometry is achieved by the
integration of a series of short-exposure images of the object
(exposure time is 5--40 ms) with the subsequent calculation of the
ensemble-averaged power spectrum of the \mbox{object
\cite{Labeyrie1970:Maximov_n},} and its phase defined from the
bispectrum \cite{Lohmann1983:Maximov_n}. In contrast to the
adaptive optics, which is effective today mainly in the infrared,
speckle interferometry can be used for observations in visible and
near UV bands. In addition, speckle interferometry is realizable
under poor atmospheric conditions, while the adaptive optics
always needs the best seeing. Another important fact is the value
of speckle interferometric equipment, which is significantly lower
than the expenses for the development of adaptive optics for a
large telescope.

\begin{figure*}
\setcaptionmargin{5mm}
\onelinecaptionstrue
\includegraphics[width=12cm, bb=0 100 780 560, clip]{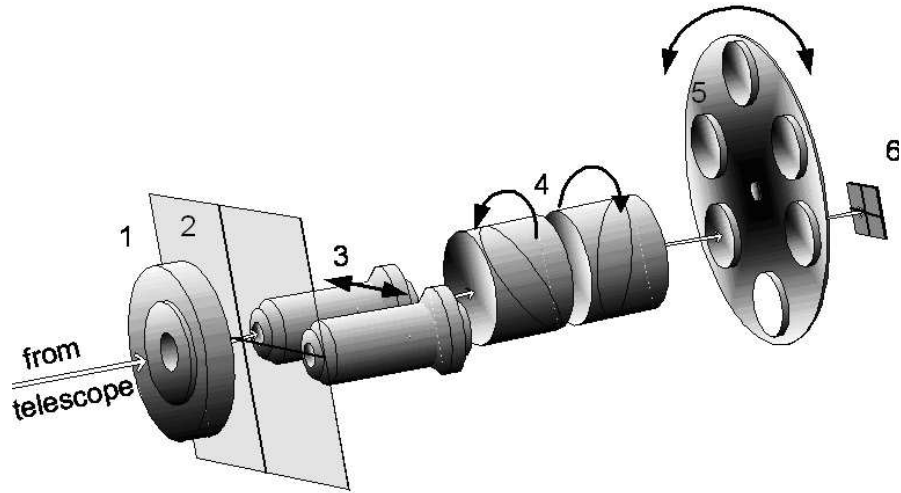}
\captionstyle{normal}
 \caption{Components of the mechanical optics unit of the BTA speckle
interferometer: 1---shutter, 3---changeable microscope objectives,
4---atmospheric chromatism compensation prisms, 5---a set of
interference filters , 6---EMCCD. Focal plane of the telescope is
enumerated 2. Black arrows point in direction of the components
movement.} \label{fig1:Maximov_n}
\end{figure*}

The speckle interferometry technique has been significantly
upgraded throughout its 35 years long history. First detectors
based on image intensifiers and film cameras
\cite{Bonneau1973:Maximov_n} were replaced by the television
photon counting systems \cite{Labeyrie1977:Maximov_n}. In the
1990s, television tubes were driven out by fast CCDs
\cite{Blazit1985:Maximov_n,Balega2002:Maximov_n}, however,
different types of image intensifiers were still used as
brightness amplifiers.  Lately, these systems were replaced by new
EMCCDs with electron charge multiplication. They have certain
advantages for application in the optical speckle interferometry:
photon counting sensitivity allied with the maximum quantum
efficiency, fast image readout, and high geometric and photometric
stability.

In the present paper we give the description of the EMCCD-based
speckle interferometer (SI), which is in active use in the
observations conducted at the BTA 6-m telescope of the Special
Astrophysical Observatory of the Russian Academy of Sciences (the
SAO RAS) since 2007. The main characteristics of the optical
components of the instrument are given in Chapter 2. Then, Chapter
3 describes the \mbox{EMCCD} detector. The control system is
briefly presented in Chapter 4. In conclusion, capabilities of the
new instrument, estimated from the observations of binary stars,
are laid out in Chapter 5. The algorithms and software for image
reconstruction from the series of short-exposure speckle
interferograms will be described in a separate paper.

\section{MAIN OPTICAL COMPONENTS}

The main optical elements of the SI optical-mechanical unit are
(Fig.~\ref{fig1:Maximov_n}): the electro-mechanical shutter with
the 6 mm aperture to block the light beam from the telescope (1),
microscope objectives for the image magnification (3), the set of
prisms for atmospheric dispersion compensation (4), a the set of
interference filters for the narrow spectral band selection (5).
The $f$:4 beam from the telescope's primary mirror forms an image
in the focal plane (2), and then it is recorded by the EMCCD
detector after the passage through all optical elements (6). Let
us consider the main components of the SI in details.

\subsection{Microscope Objectives}

The SI microscope objectives are designed for matching the image
scale in the prime focus of the BTA with the detector's pixel
size. The size of a single speckle in a speckle interferogram,
corresponding to the diameter of the first dark ring in the Airy
disk at the wavelength  $\lambda$, is equal to $d=1.22\lambda/D$,
where $D$ is the aperture diameter. In the BTA focal plane, for
$\lambda_{0}$=550~nm $d$=2~$\mu$m. The detector pixel size is 16
$\mu$m, therefore, following the Kotelnikov's theorem
($f_{discr}\geq 2f_{max}$, where $f_{max}$ is the maximum
frequency in the signal spectra), an image magnification of
approximately 20 times must be applied to reconstruct the speckle
profile.

\begin{figure*}
\setcaptionmargin{5mm}
\onelinecaptionstrue
\includegraphics[width=12cm, bb=0 0 480 280, clip]{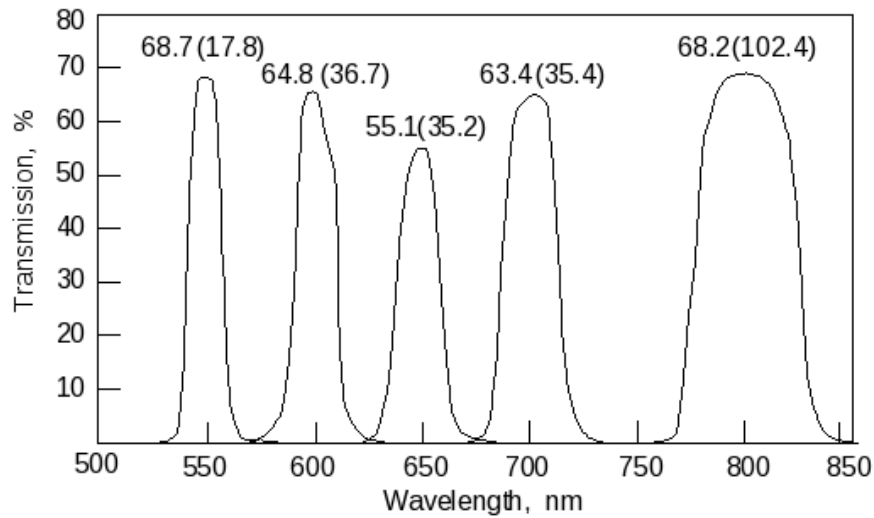}
\captionstyle{normal} \caption{ Spectral characteristics of
interference filters. The transmission in maximum and half widths
(in brackets) are marked over each curve. } \label{fig2:Maximov_n}
\end{figure*}

We use two high-quality Carl Zeiss objectives with the corrected
field curvature to match the scales: a 2.5-fold achromat with the
numerical aperture 0.08 and a 16-fold plan-achromat with the
numerical aperture 0.35. They both have a back working distance of
160 mm and form a magnified, inversed and free from aberrations
image of the star at the CCD.  With the BTA main mirror curvature
radius of $R$=48051$\pm$15 mm (floor measurement by L.I.~Snezhko
using the Hartmann method), the microscope objectives provide
scales of 3.44$''$/mm and 0.54$''$/mm on the detector. For the
512$\times$512 pixel detector with the size of the sensitive area
8.19$\times$8.19 mm, the corresponding scales are 0.0550$''$/pix
and 0.0087$''$/pix, giving the fields of view of 28.2$''$ and
4.4$''$. The 16-fold objective is the main working objective of
the SI, while the small magnification is used only for pointing on
the object and tying-in the position angles of the system.

For some purposes that require obtaining a detailed profile of the
Airy disk, one might need to use higher optical magnifications.
One of the examples is the image reconstruction of visible disks
of cool supergiant stars. For such cases, we anticipated an
installation of a 32-fold microscope objective with a
corresponding reduction of the field of view.

Small convergence of the beam after the 16-fold objective (1:64)
permits fitting other optical elements (prisms, filters) on the
axis between the objective and the detector, without resorting to
additional transfer optics.

\subsection{Interference Filters}

To keep the coherence during the interferometric picture
recording, the width of the filter transmission band
$\Delta\lambda$ is selected following the relation

\begin{equation}
\label{formula1:Maximov_n}
 \frac{\Delta\lambda}{\lambda}<\frac{r_{0}}{D},
\end{equation}

\noindent where $\lambda$ is the center wavelength of the filter
transmission band, $r_{0}$ is the atmospheric coherence radius
(Fried parameter) \cite{Fried1965:Maximov_n}, or

\begin{equation}
\label{formula2:Maximov_n}
  \frac{\Delta\lambda}{\lambda}<\frac{2\pi}{\sigma_{r}},
\end{equation}

\noindent where $\sigma_{r}$ is the mean-square aberration of the
incoming wavefront phase given in radians.

Under the seeing of approximately 1$''$, ($r_{0}$$\approx$20 cm),
the filter limitations for the BTA applications are chosen by the
ratio $\Delta\lambda/\lambda\approx$1/30. For poor atmospheric
conditions the filter passband must be reduced.  However, in most
of the cases the filter transmission band can be doubled without
significant dissipation of the high-resolution data. This
simplification leads to a larger detected flux. In the near
infrared, the filters with a band width of 100 nm can be used.
Narrower filters can be installed in the beam to cut out the
selected bands in the spectrum of an object. One of the examples
of such tasks is the Mira stars image reconstruction in the TiO
absorption bands and in the nearby continuum.

\begin{figure}
\setcaptionmargin{5mm}
\onelinecaptionstrue
\includegraphics[width=8cm, bb= 30 35 463 437, clip]{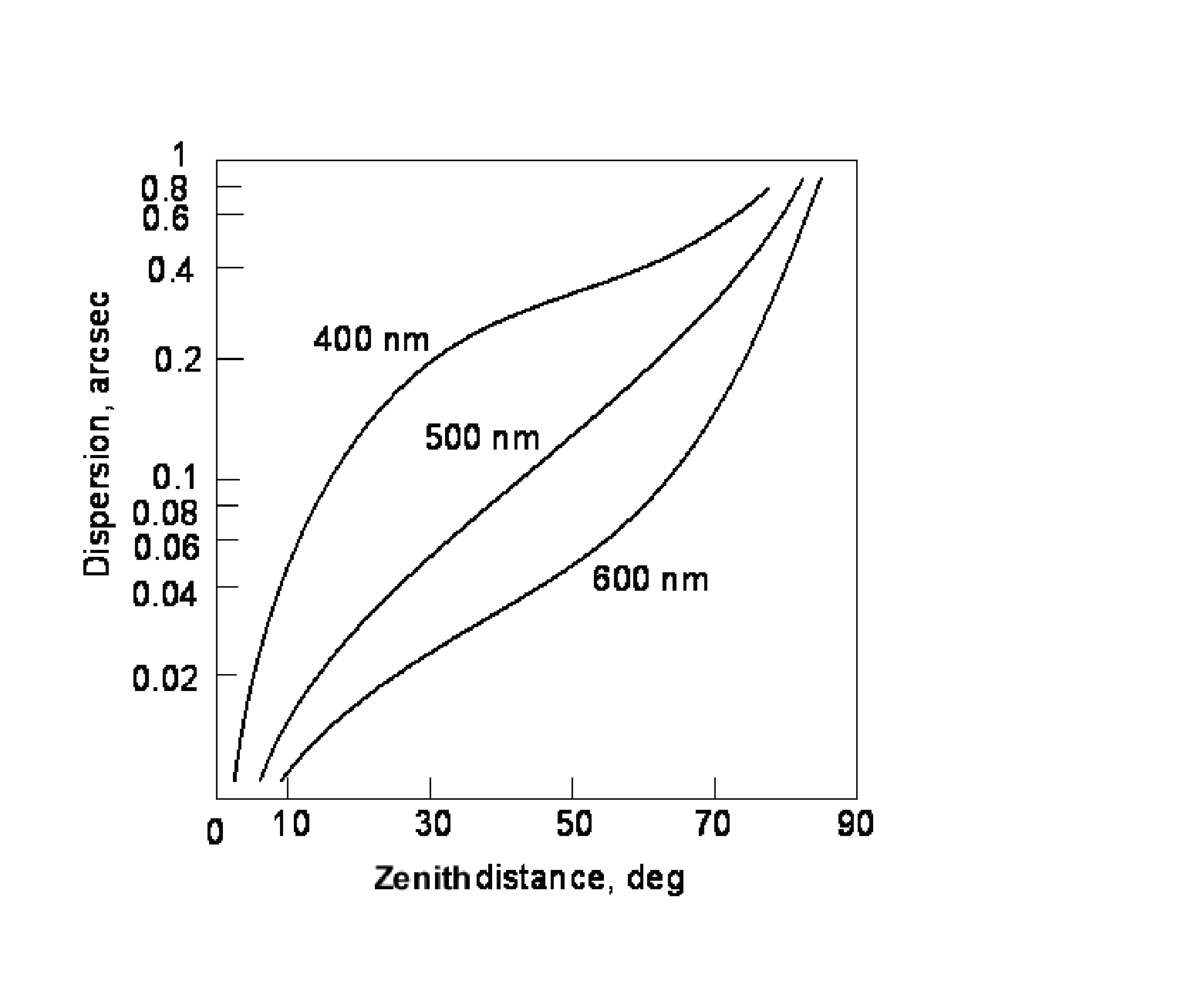}
\captionstyle{normal} \caption{ Dependance of the atmospheric
dispersion from the zenith distance for three wavelengths. }
\label{fig3:Maximov_n}
\end{figure}

Five interference filters, manufactured by Andover (USA), are
arranged on the turret assembly, and are being brought in the beam
by the rotating mechanism. All the filters have the diameter of 25
mm (the light diameter is 21.2 mm) and thickness of \mbox{6 mm.}
The parameters of the filters are given in
Fig.~\ref{fig2:Maximov_n}. The thickness of the filter dielectric
layers is constant (0.25\%) throughout the surface in order to
ensure the design performance. During the detector calibration
procedures or the system testing in the white light, no filter is
needed in the optical path. For these purposes the filter wheel
has one additional hole.

\subsection{Atmospheric Dispersion Correction}

The atmospheric dispersion has an effect on the resolution in
speckle interferometry even while applying narrow interference
bands. The dispersion value can be estimated from the empirical
formula proposed by Lambert as far back as in 1759:

\begin{equation}
\label{formula3:Maximov_n}
 \frac{dz}{d\lambda}=-\frac{\tan z}{n_{a}}\frac{d n_{a}}{d\lambda},
\end{equation}

\noindent where $n_{a}$ is the refraction index of the atmosphere,
$z$ is the zenith distance of the object.

The Owens empirical expression \cite{Owens1967:Maximov_n} is most
frequently used to describe the wavelength dependence of the
refraction index

{\setlength\arraycolsep{2pt}
\begin{eqnarray}
\label{formula4:Maximov_n} \nonumber (n_{a}-1)10^{8}  =
\biggl[2371.34+\frac{683939.7}{(130-\lambda^{-2})}\\
+\frac{4547.3}{(38.9-\lambda^{-2})}\biggr] D_{s}
+\biggl[6487.31+58.058\lambda^{-2}\\ \nonumber
-0.71150\lambda^{-4}+0.08851\lambda^{-6}\biggr]  D_{w},
\end{eqnarray}}

\noindent where $D_{s}$ and $D_{w}$ are the factors accounting for
the density of the dry air and water vapors, respectively.  The
value of $n_{a}(\lambda)$ can be estimated with an accuracy of
10$^{-7}$--10$^{-8}$ from formula (\ref{formula4:Maximov_n}).
Fig.~\ref{fig3:Maximov_n} shows the dependence of zenith distance
from the refraction index in three wavelengths for the BTA site
(pressure 720 mbar, temperature 0$\degr$C, relative humidity 30\%)
as follows from expressions ~(\ref{formula3:Maximov_n}) and
(\ref{formula4:Maximov_n}).

For example, an image of a point source at $z$=30$\degr$ and
$\lambda$=500 nm will be shifted by 0.05$''$ from its true
position. The vertical extension of speckles in the region
$\lambda$=500 nm through the $\Delta\lambda$=20 nm filter will be
around 0.02$''$, which is comparable with the size of the
diffraction spot.

The simplest way to compensate the atmospheric chromatism is the
use of a direct-vision prism with the dispersion vector opposite
to the atmospheric dispersion. The angular dispersion of the prism
must change with the zenith angle. One should keep in mind that in
the SI we use the 16-fold magnification, therefore the prism
corrector, which is installed behind the microscope objective,
must provide a proportionally higher dispersion.

Beam deviation $\alpha$ in the prism with the index of refraction
$n$ and the apex angle $\beta$ is equal to

\begin{equation}
\label{formula5:Maximov_n}
 \alpha=(n(\lambda)-1)\beta.
\end{equation}

For a prism consisting of a pair of wedges with the angles
$\beta_{1}$ and $\beta_{2}$ and the indices of refraction $n_{1}$
and $n_{2}$, the refracting angle is

\begin{equation}
\label{formula6:Maximov_n}
 (n_{1}(\lambda)-1)\beta_{1}-(n_{2}(\lambda)-1)\beta_{2}.
\end{equation}

The wedge angles ratio is obtained by specifying the zero
deviation at $\lambda_{0}$=550 nm:

\begin{equation}
\label{formula7:Maximov_n}
 \frac{\beta_{2}}{\beta_{1}}=\frac{n_{1}(\lambda_{0})-1}{n_{2}(\lambda_{0})-1}.
\end{equation}

To compensate the atmospheric dispersion at the maximum zenith
angle $z$=60$\degr$ in the 20 nm band at $\lambda_{0}$ =550 nm,
the following equation should be valid:

\begin{equation}
\label{formula8:Maximov_n}
 \frac{dz}{d\lambda}=\beta_{1}\left( \frac{d n_{1}}{d\lambda}\right)_{\lambda_{0}}-
\beta_{2}\left( \frac{d n_{2}}{d\lambda}\right)_{\lambda_{0}},
\end{equation}

\noindent leading to the apex angle of the first wedge:

\begin{equation}
\label{formula9:Maximov_n}
 \beta_{1}=\frac{\Delta z}{\Delta n_{1} - \Delta n_{2} \frac{n_{1}(\lambda_{0})-1}{n_{2}(\lambda_{0})-1}},
\end{equation}

\begin{figure}
\setcaptionmargin{5mm}
\onelinecaptionstrue
\includegraphics[width=7.8cm, bb= 33 35 457 436, clip]{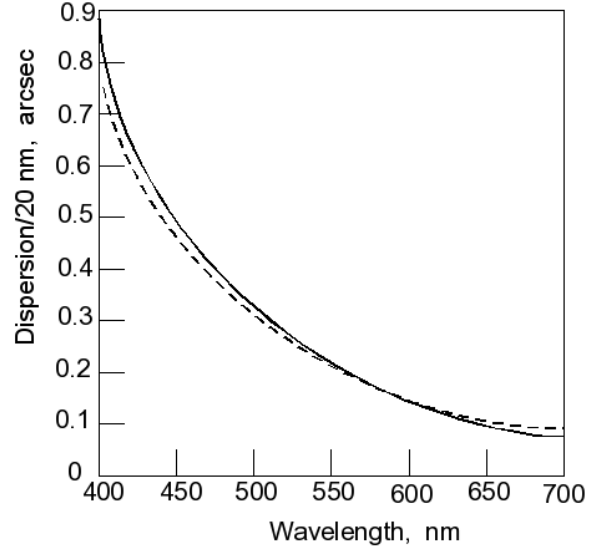}
\captionstyle{normal} \caption{ Dispersion curves of the prism
(solid line) and the atmosphere (dashed line). }
\label{fig4:Maximov_n}
\end{figure}

\noindent where $\Delta$ means the variation range of $z$ and $n$ in
the wavelength range from $\lambda_{0}-10$ nm to $\lambda_{0}+10$
nm. To correct the atmospheric chromatism at smaller $z$, the
compensating prism should provide lower dispersion.

In our observations until September 2007, a direct-vision Amici
prism from a pair of wedges has been used for dispersion
compensation. The dispersion value has been changed by translation
of the prism along the axis: the lowest value was obtained in the
immediate proximity of the detector; the highest dispersion was
reached at the remote position. The advantage of such construction
consists in its low light losses in the glass and on the surfaces.
The disadvantage of the assembly is its remaining dispersion in
the circumzenithal zone. This dispersion cannot be eliminated as
the prism cannot be installed closer than some minimum distance
from the image acquisition plane.

The dispersion curve for the prism compared to the atmospheric
dispersion is shown in Fig.~\ref{fig4:Maximov_n}. It can be seen
that by the selection of the wedge angles and refractive indices,
it is possible to match the dispersion variation of the prism and
atmosphere in the range of 550--700 nm.

The new compensator was built on the base of a direct-vision
Risley prism consisting of a pair of identical prisms rotating
around the instrument axis in opposite directions (see
Fig.~\ref{fig1:Maximov_n}). The compensator gives a dispersion of
$2\gamma\sin(\theta/2)$, where $\gamma$ is the dispersion of a
single prism, $\theta$ is the rotation angle of the prisms. Each
prism in the assembly is a glued pair of wedges (crown + flint)
with carefully selected indices of refraction, providing a
dispersion of 330$''$/20 nm at the direct-vision wavelength of 550
nm. Shorter and longer wavelengths deviate symmetrically in
opposite directions from the 550 nm beam. The net deviation of the
beam is the vector sum of the two separate deviation vectors. If
the prisms are rotated by 90$\degr$, they combine to act as a
parallel plate with no net angular deviation. To simplify the
construction, the rotation mechanism of the prisms is made
dependent, providing the rotation of the prisms by equal angles
about the optical axis. The rotation angle of the prisms is
defined by the computer when the object is chosen and its zenith
distance is calculated. The calibration of the $\theta(z)$
relationship is made using the observations of bright stars at
different $z$ during the nights of good seeing.

Due to different distances of the prisms from the detector plane
(the front prism is at larger distance and produces larger
dispersion), the net deviation of the assembly is shifted from the
vertical. The value of this shift is zenith angle dependent. The
remaining residual dispersion can be easily compensated via
rotating the prism mounting by a small angle \mbox{(about
10$\degr$)} in the opposite to the differential dispersion
direction.

The Risley compensator, consisting of a pair of rotatable wedged
elements, provides the correction of the atmospheric chromatism of
speckle interferograms in the range of zenith distances from
0$\degr$ to 60$\degr$. The prisms were manufactured by B.Halle
Nachfolger GmbH, Germany, from the SF L6 crown glass and heavy
flint glass LaSF N30. The apex angles of the wedges made of crown
glass and heavy flint glass are $\beta_{1}$=41.6$\degr$ and
$\beta_{2}$=41.9$\degr$, respectively. The prisms have a 5-layer
antireflection coating for the range of 450--1100 nm.

\section{DETECTOR}

\subsection{General Characteristics}

\begin{figure*}
\setcaptionmargin{5mm}
\onelinecaptionstrue
\includegraphics[width=7.8cm]{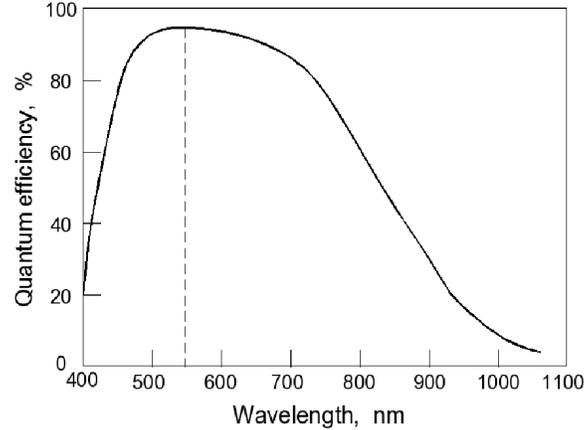}
\captionstyle{normal} \caption{ Rated curve of the CCD97 quantum
efficiency. Dashed line marks the wavelength corresponding to the
maximum quantum efficiency of 93\%. } \label{fig5:Maximov_n}
\end{figure*}

The PhotonMAX-512B camera developed by Princeton Instruments, USA,
on the base of electron multiplication CCD97 (EMCCD) is used as
the detector in the SI. The back-illuminated CCD has 16$\times$16
$\mu$m pixels in a 512$\times$512 frame transfer format with the
total photosensitive area of 8.19$\times$8.19 mm. The quantum
efficiency of the CCD is QE=93\% at 550 nm. The quantum efficiency
curve at different wavelengths is shown in
Fig.~\ref{fig5:Maximov_n}.

\begin{figure}
\setcaptionmargin{5mm}
\onelinecaptionstrue
\includegraphics[width=7.8cm, bb=0 0 450 380, clip]{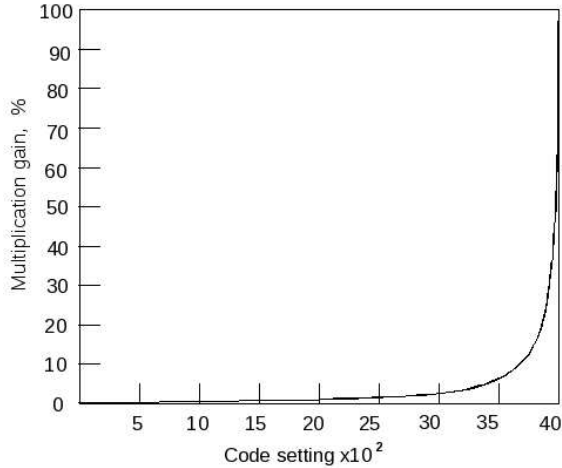}
\captionstyle{normal} \caption{ CCD97 charge multiplication factor
dependence on the code setting at the input of digital-to-analog
converter. } \label{fig6:Maximov_n}
\end{figure}

The EMCCD has two independent readout registers on the output. The
first one is used for standard, high dynamic range application as
in traditional CCDs, the second one is devoted for high-speed, low
light level multiplication gain mode. In the normal readout
application (no multiplication), the readout mean-square noise is
45 $e^{-}$ and 60 $e^{-}$ at operating readout frequencies 5 and
10 MHz, respectively. Under maximum multiplication of the charge
generated in the pixels, the readout noise falls \mbox{below 1
$e^{-}$.} The PhotonMAX can offer a multiplication gain factor
between $\times1$ and $\times1000$. The typical relationship
between the 12-bit setting of the DAC (digital-to-analog
converter), which defines the voltage applied to multiplication
register clocks, and the charge multiplication factor is close to
exponential (see Fig.~\ref{fig6:Maximov_n}). The maximum
multiplication is achieved under the voltage of about 43\,V on the
extended multiplication register. Further register voltage gain
can destroy the CCD. Most of the time we use the PhotonMAX-512B
camera in the high-multiplication mode, and the normal mode (no
gain) with low multiplication is used only during the speckle
observations of bright stars.

To reduce the dark current, the EMCCD is cooled by a 4-stage
thermoelectric Peltier cooler. It is stabilized to within
$\pm$0.05 $\degr$C ~by a feedback control system. The lowest
achievable temperature of the CCD is --70$\degr$C providing the
dark current below 0.01 $e^{-}$/s per pixel.

The CCD is mounted on the cold finger in the vacuum module. The
photosensitive area of the CCD is protected by a single optical
entrance window, which provides low light losses. The heat
produced by the Peltier device is removed from the camera by an
inbuilt fan. The fan is designed for low vibration and does not
adversely affect the image.

The camera can be used in the frame transfer mode in the wide
range of exposures between 10 $\mu$s to 10 min. Maximum readout
rate of the camera in the full-frame mode is 29 frames per second
(fps). As the PhotonMAX camera uses a frame-transfer CCD, it does
not incorporate any electro-mechanical shutters that set the
exposure time. The exposure time is set electronically by the
software. The external shutter 1 (see Fig.~\ref{fig1:Maximov_n})
is used solely to block the incoming light in the SI.

\subsection{Time Stability}

Basic properties of the PhotonMAX camera were studied at the
laboratory and on the telescope. The study included measurements
of the wavelength dependence of the quantum efficiency of the SI,
an analysis of the system's stability, its frame rate in different
acquisition modes, the signal homogeneity over the field, and the
flat field response of the system.

\begin{figure*}
\setcaptionmargin{5mm}
\onelinecaptionstrue
\includegraphics[width=11cm, bb=15 0 490 168, clip]{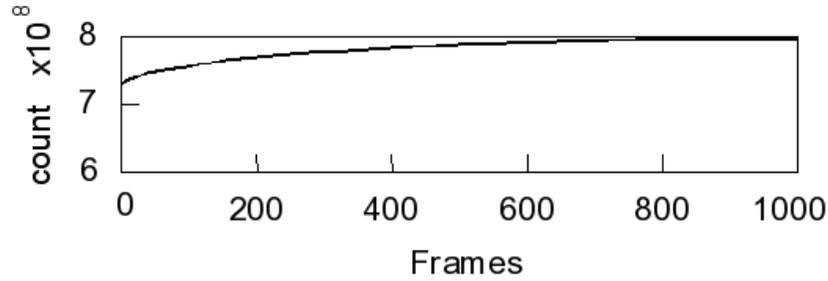}
\captionstyle{normal} \caption{ CCD flat field time stability. }
\label{fig7:Maximov_n}
\end{figure*}

Speckle interferometry operates with series of a few thousand
short-exposure images of an object. The total integration time of
the series is less than a few minutes as the optical transfer
function of the atmosphere is not stable for longer periods. To
check the time stability of the detector, we have accumulated a
few sets of 1000 uniform (flat) light fields obtained at 10 MHz
operating frequency with an exposure time of 10 ms. In
Fig.~\ref{fig7:Maximov_n} presents the variations of the summed
signal from the CCD. It can be seen that the CCD signal intensity
varies exponentially by almost 10\% during the
acquisition of 1000 frames (54 s). This circumstance must be taken
into account for photometric measurements. To stabilize the
multiplication level of the CCD before the main exposure, the CCD
has to be read for more than 5 min.

\subsection{Readout Speed}

\begin{figure*}
\setcaptionmargin{5mm}
\onelinecaptionstrue
\includegraphics[width=11cm]{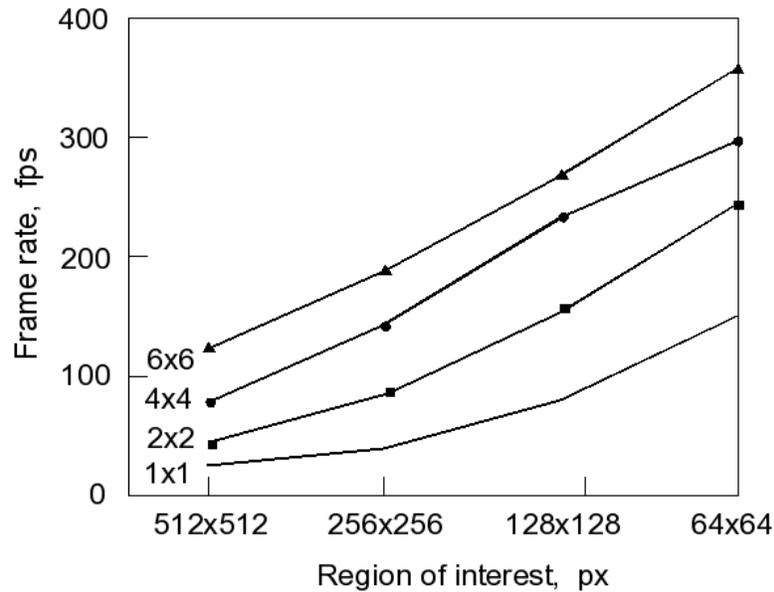}
\captionstyle{normal} \caption{ Frame rate dependence of the CCD
reading area without binning, and with 2$\times$2, 4$\times$4 and
6$\times$6 binning. The dimensions of the reading zone area are
protracted on the X-axis. } \label{fig8:Maximov_n}
\end{figure*}

Regardless of the exposure time, the highest readout speed of the
512$\times$512 pix full format image is 29 fps. Binning (combining
pixels into one super pixel) and sub-region selection allow
increasing the sensitivity and frame rate of the PhotonMAX-512B.
The camera allows the binning of 2$\times$2, 4$\times$4 and
6$\times$6 along either direction of the CCD. The maximum frame
rate vs. binning for different readout regions is shown in
Fig.~\ref{fig8:Maximov_n}. The maximum rate of 350 fps is obtained
for 6$\times$6 binning and reading only the 1/64 area of the CCD.
However, all types of binning are rarely used in speckle
interferometry as it reduces spatial resolution.

\subsection{Flat Field Signal and the Dark-Charge Pattern}

Even under uniform incoming illumination (flat field) of the CCD,
the obtained image is not perfectly flat. Inhomogeneity of the
response is explained by different reasons:

\begin{itemize}
\item[-] lack of uniformity of the chip; \item[-] contamination
and granularity of the camera's entrance window; \item[-]
imperfections of the CCD readout electronics, which become
apparent mostly on the edge zones of the frame.
\end{itemize}

Flat field problems do not play such an important role in speckle
image reduction as compared to the classical long-exposure
astronomical imaging due to the fact that interferometry exploits
ensemble-averaged power spectra. Nevertheless, the standard
reduction procedure in speckle interferometry also implies a
flat-field reduction of each speckle interferogram.

\begin{figure*}
\setcaptionmargin{5mm}
\onelinecaptionstrue
\includegraphics[width=12cm]{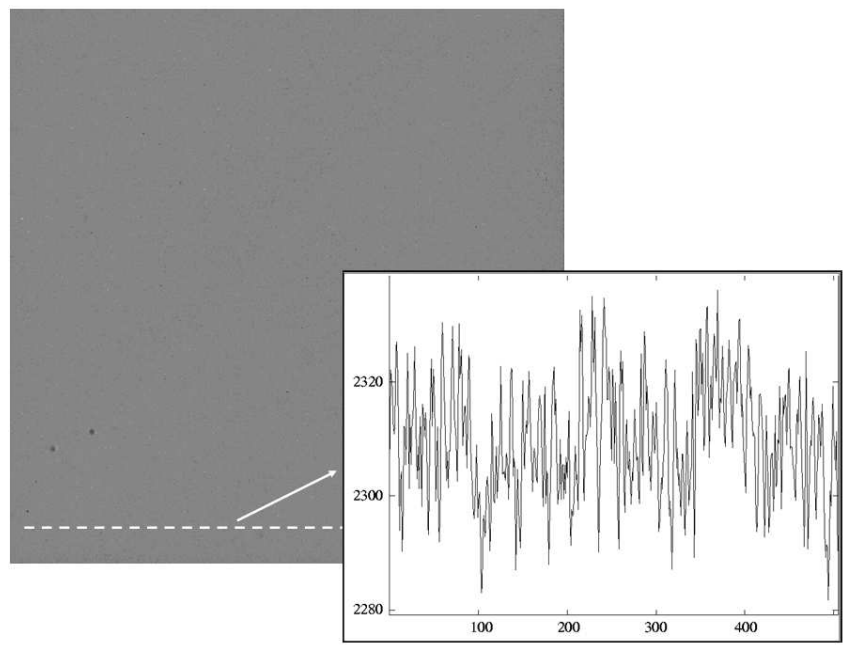}
\captionstyle{normal} \caption{ Image of a small area of the
twilight sky integrated in sets of 2000 frames (a flat field). A
cross section along one of the lines in the bottom part of the
frame is shown on the right. } \label{fig9:Maximov_n}
\end{figure*}

We have studied the SI flat field by integration of the twilight
sky. One of the examples obtained by averaging of 2000 short
exposures is presented in Fig.~\ref{fig9:Maximov_n}. The field
inhomogeneity (signal dispersion) for this image is 0.5\%. Summing
a larger number of images allows determining the PhotonMAX-512B
flat filed with the noise dispersion under 0.05\%.

\begin{figure*}
\setcaptionmargin{5mm}
\onelinecaptionstrue
\includegraphics[width=12cm]{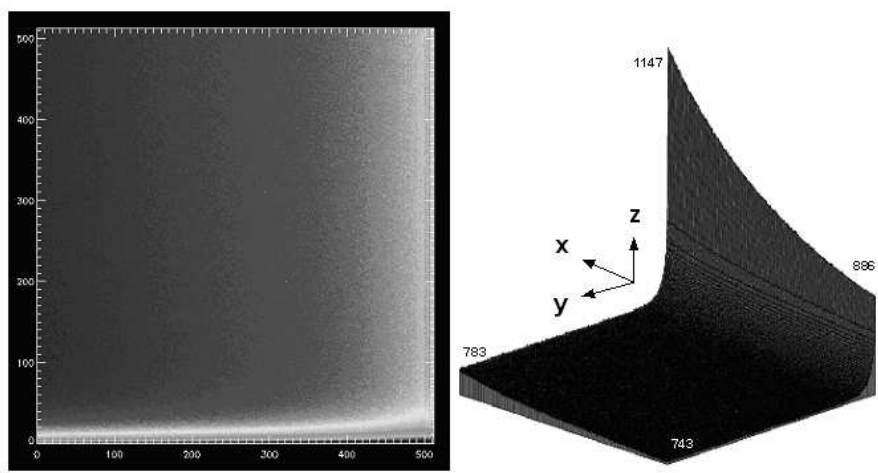}
\captionstyle{normal} \caption{ View of the CCD background and its
3-D presentation. } \label{fig10:Maximov_n}
\end{figure*}

The CCD offset signal measurements with no light coming into the
camera (bias) were performed with the closed shutter using 10
$\mu$s exposures. An average of 500 frames obtained with the
highest CCD multiplication is shown in Fig.~\ref{fig10:Maximov_n}.
The general view of the base plate is shown on the left, and its
3-dimensional presentation is on the right. Pixel values of the
charge are given in the corners of the bit map. It can be seen
that first lines in the frame have significantly higher charges
due to the charge inleakage from the chip edges. Maximum
difference in the counts is 0.6\% from the full range of the
analog-to-digital converter scale (65536). The bias pattern is
fully subtractable if the dark-charge pattern is acquired under
conditions identical to those used to collect the actual speckle
images.

\subsection{Single Photon Sensitivity}

\begin{figure*}
\setcaptionmargin{5mm}
\onelinecaptionstrue
\includegraphics[width=12cm]{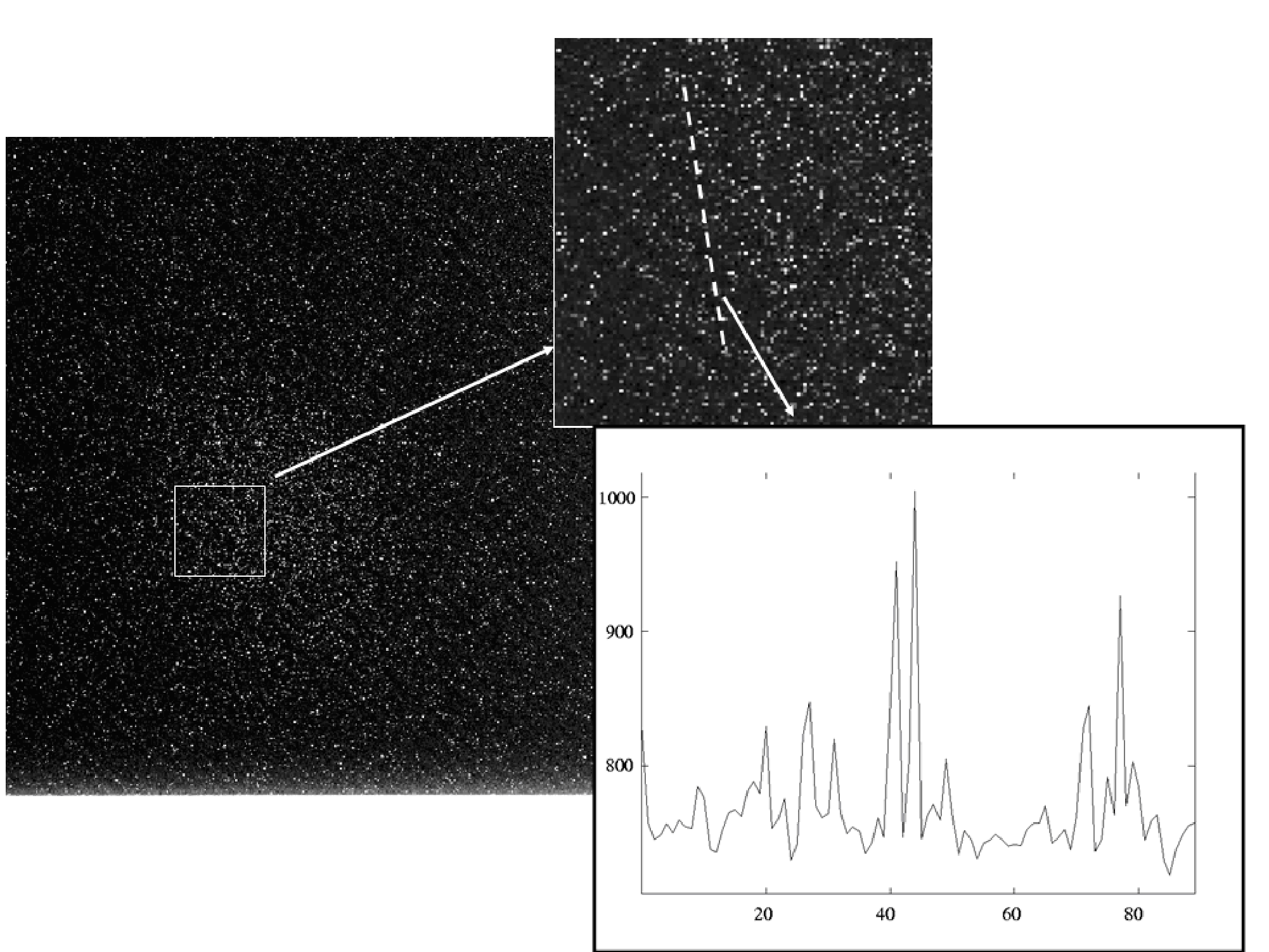}
\captionstyle{normal} \caption{ Left: a speckle image of the star
LkHA~198, acquired by the PhotonMAX-512B in the maximal CCD signal
multiplication mode. Top right: a magnified segment with the
maximal flux level. Bottom right: an arbitrary cross-section,
demonstrating the amplitudes of independent photon events. }
\label{fig11:Maximov_n}
\end{figure*}

During the speckle observations of faint stars with the maximum
gain, the PhotonMAX-512B acquires images in the photon counting
mode. Fig.~\ref{fig11:Maximov_n} shows the 20 ms exposure image of
LkHA 198 ($m_{V}$=14.3$^m$) obtained at the 6-m telescope in the
600/40 nm filter under $1''$ seeing. The cross-section illustrates
intensity maxima corresponding to single photon events. The CCD
noise dispersion for this image is 30--40 counts. Individual
photons are detected by the PhotonMAX-512B with a signal-to-noise
ratio \mbox{of 10 to 50.}

\section{CONTROL SYSTEM}

\begin{figure*}
\setcaptionmargin{5mm}
\onelinecaptionstrue
\includegraphics[width=13cm, bb=10 46 622 448, clip]{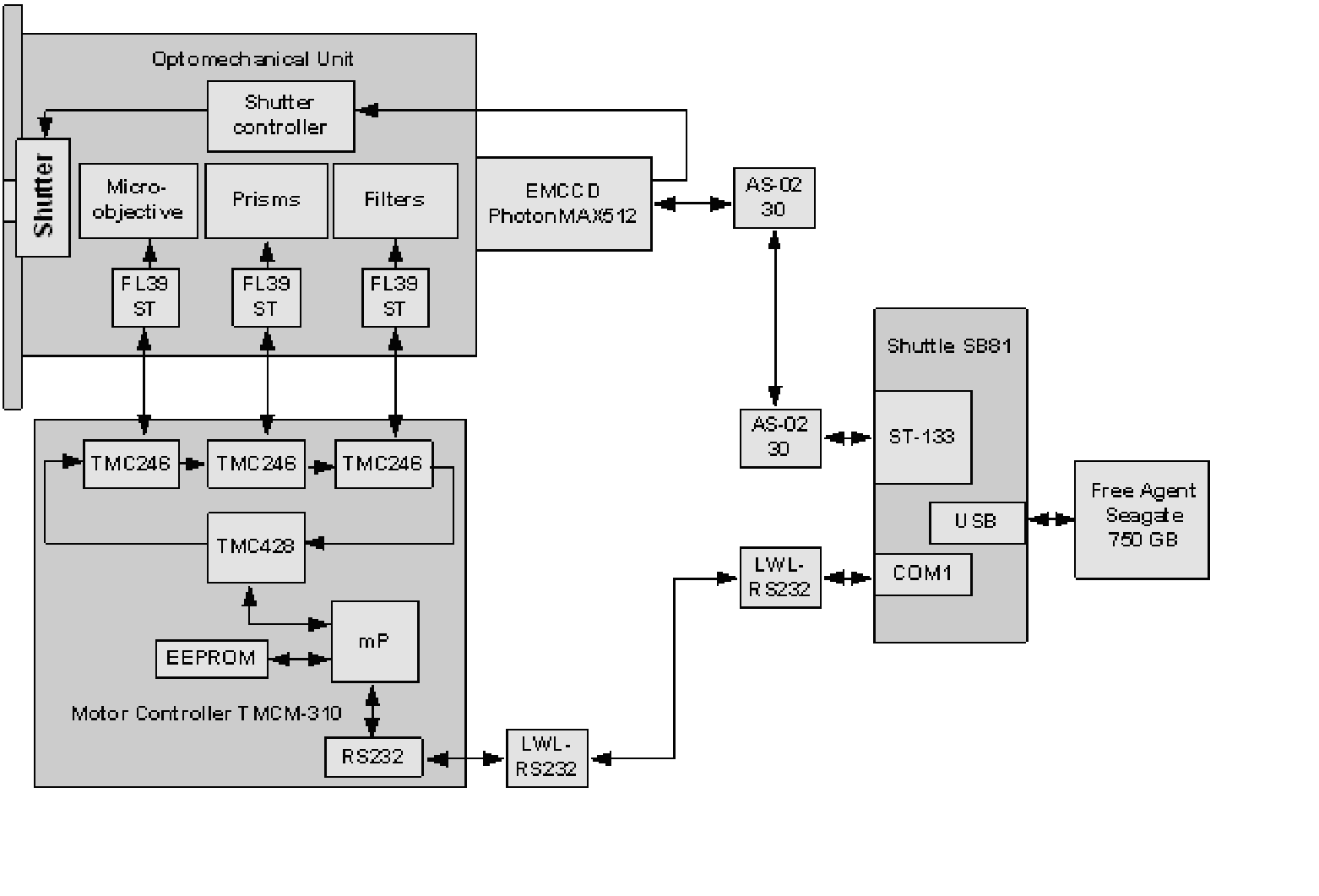}
\captionstyle{normal} \caption{ Block diagram of the SI control
system. } \label{fig12:Maximov_n}
\end{figure*}

The SI control system diagram is given in
Fig.~\ref{fig12:Maximov_n}. Control of optical elements is
performed by the stepper motor drive controller TMCM-310 from
Trinamic Motion Control GmbH \& Co., Germany. It can drive up to
three bipolar stepper motors through the TMC246 drivers with a
peak coil current of up to 1.5\,A. A built-in microprocessor
($\mu$P) and a nonvolatile 16 KB memory (EEPROM) with the aid of
the parameter setting module TMC428 allow flexible adjustment of
the drive for selected translation mechanisms. The controller is
plugged in the COM-port of the control computer through the
standard series RS-232 interface via the 8-channel fiber optic LWL
converters.

The microscope objectives are translated by the precision linear
motorized translation stage ERLIC 85 produced by OWIS GmbH,
Germany. The setting accuracy is 2 $\mu$m for the 50 mm travel.

The filter wheel is fixed on the rotation stage mechanism DMT-65.
A filter can be installed in the beam with the positional accuracy
of 0.01$\degr$. The atmospheric dispersion compensation prisms are
rotated by stepper motors through a differential mechanism.
Special guidance provides a practically slip-stick free movement
as well as high load capacity for all translation stages.

The camera and other units of the SI are controlled by one
computer based on Intel Pentium 4 3.2GHz processor. We use the
Shuttle XPC SB81P Barebone system with low electromagnetic
radiation. The PhotonMAX-512B camera control is conducted by the
ST-133 controller through a PCI-bus via the fiber optic AS-0230
receiver-transmitters. RAM is increased to 2 GB to enable rapid
accumulation of up to 1940 speckle images, which are then recorded
on the computer hard drives in the binary files of SPE type with a
4100 byte header. The file header contains all the important
camera settings that were used during the accumulation of the
speckle images. The accumulation of the data can be performed on
the high-capacity external hard drives via a USB-interface.

 \begin{figure*}
\setcaptionmargin{5mm}
\onelinecaptionstrue
\includegraphics[width=12.5cm]{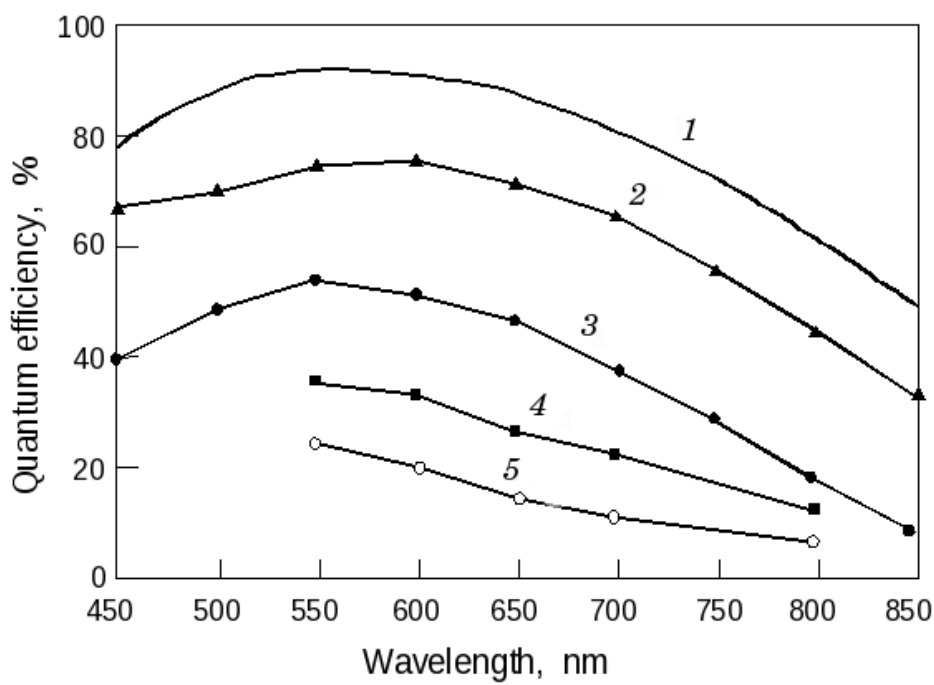}
\captionstyle{normal} \caption{ SI quantum efficiency corrected
for light losses on the optical elements: 1---CCD97, 2---CCD97 +
PhotonMAX-512B entrance window, 3---CCD97 + PhotonMAX-512B
entrance window + dispersion compensation prism, 4---CCD97 +
PhotonMAX-512B entrance window + dispersion compensation prism +
filters, 5---total efficiency corrected for losses on the
miroscope objectives.} \label{fig13:Maximov_n}
\end{figure*}

During the observations we often need to monitor the accumulation
of power spectrum of the object in the real time mode. The
information on duplicity (multiplicity) of the object at the
initial stage allows assessing the appropriateness of further
acquisition. To solve this problem we are currently implementing
reconfigurable multiprocessor computing systems into the process
of observations and data reduction. Unlike systems based on rigid
architecture, reconfigurable systems allow on-the-fly changes of
the calculation parameters. As a result, users may fit the
system's architecture to the structure of a given task.

\begin{figure*}
\setcaptionmargin{5mm}
\onelinecaptionstrue
\includegraphics[width=12cm, height=6cm, bb=15 140 975 630, clip]{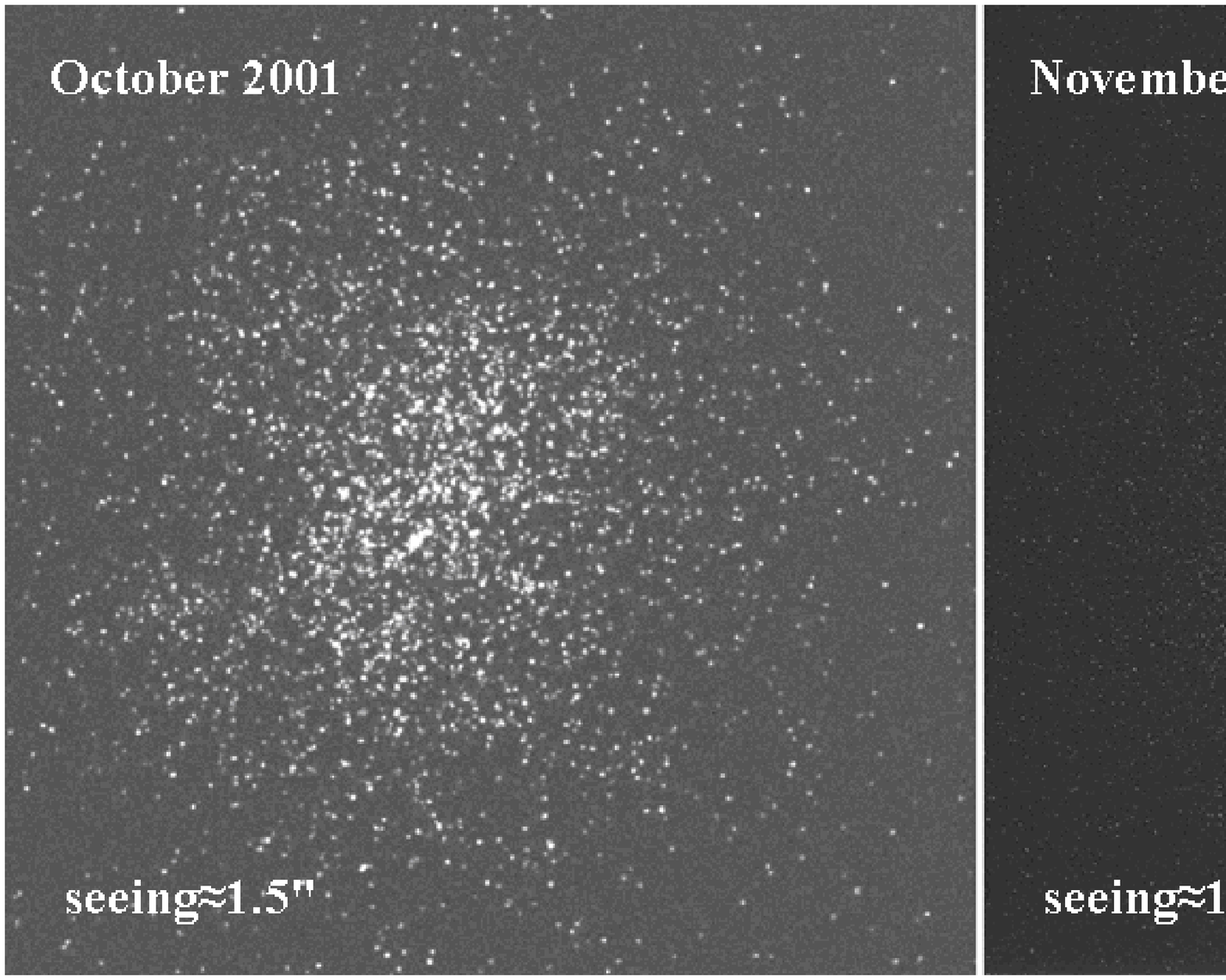}
\captionstyle{normal} \caption{ Individual speckle images of
DF~Tau binary star. Left: an image, acquired with the CCD + image
intensifier system. Right: image obtained with the EMCCD. }
\label{fig14:Maximov_n}
\end{figure*}

\begin{figure*}
\setcaptionmargin{5mm}
\onelinecaptionstrue
\includegraphics[width=12cm]{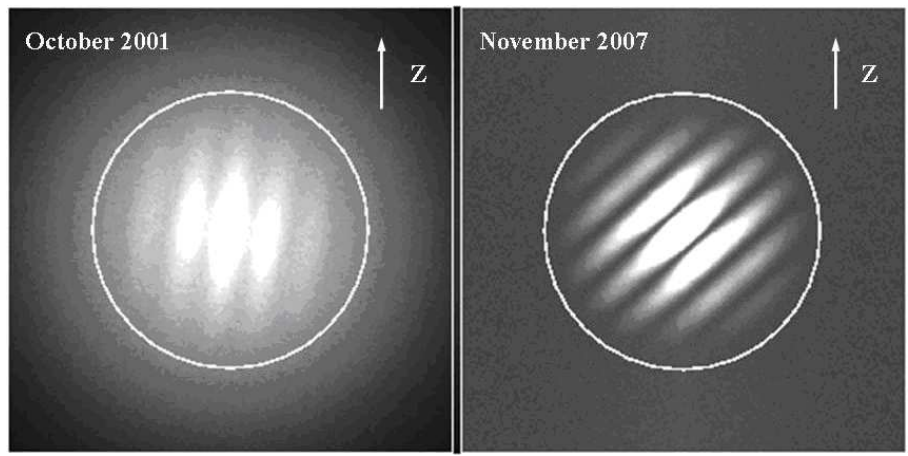}
\captionstyle{normal} \caption{ DF~Tau power spectra. Left: the
spectrum accumulated on 2000 speckle interferograms with the CCD +
image intensifier system. Right: the same accumulation with the
EMCCD. The arrow points at the direction of zenith. }
\label{fig15:Maximov_n}
\end{figure*}

To calculate the power spectrum in real time mode, a unit based on
a reconfigurable accelerator of personal computers (RAPC-25) with
25 gigaflops processing power \cite{Kalyaev2008:Maximov_n} was
designed in the Research Institute of Multiprocessor Computing
Systems, Taganrog, Russia. This unit implements parallel
algorithms to correct the optical images for atmospheric
distortion, allows real-time reduction of speckle interferograms,
and on-the-fly calculations of the power spectrum of the observed
object's image. The unit is reprogrammable, an option that can be
used, e.g., for object phase reconstruction.

\section{EFFICIENCY OF THE SYSTEM}

Quantum efficiency of the SI was measured in 9 wavelengths using
the MDR-41 grating monochromator, and the OL 730D programmable
DSP-radiometer, which is sensitive in the range of \linebreak
200--1000 nm. Fig.~\ref{fig13:Maximov_n} shows the system's
efficiency curves vs. wavelength, adjusted for losses at the
optical components: the upper curve (1) gives the QE of the CCD97
chip, the lowest curve (5) presents the total QE of the SI. Note
that the CCD itself has a 20\% efficiency at $\lambda$=950 nm.

We compared the PhotonMAX-512B camera performance with the
previous detector, which has been used for speckle observations
until 2007. This camera was built on the base of a 3-stage
electrostatic-focusing image intensifier AEG1510 (Siemens,
Germany) optically coupled to the fast CCD camera \mbox{SensiCam}
370LL, manufactured by the PCO Computer Optics, Germany.

Individual speckle interferograms of the binary star DF Tau
($m_{V}$=12$^{m}$) obtained with the 6-m telescope in October 2001
(intensifier--SensiCam) and November 2007 (PhotonMAX-512B) are
shown in Fig.~\ref{fig14:Maximov_n}. Both observations were
performed under \mbox{1$''$--1.5$''$} seeing in the
$\lambda/\Delta\lambda$=800/100 nm filter. In 2001, a scale of
0.004$''$/pix was provided by the 32$\times$ microscope objective,
while in 2007 a scale of 0.0087$''$/pix was provided by the
16$\times$ image magnification. The difference between the two
images is evident. Photon noise is dominating at the
intensifier--SensiCam speckle image. To reconstruct an image of
the binary from such data, it is necessary to process up to
10$^{4}$ individual speckle interferograms. On the other side, the
speckle structure is clearly seen on the PhotonMAX-512B image; in
this case a diffraction-limited reconstruction of the DF Tau image
can be obtained from several tens of frames.

In speckle interferometry, the power spectrum of the star's image
is recovered from a series of short-exposure speckle
interferograms using the Labeyrie's method
\cite{Labeyrie1970:Maximov_n}. For binary and multiple star
speckle interferometry at the BTA telescope we use the procedures
described in \cite{Balega2002:Maximov_n,Maximov2003:Maximov_n}.
Fig.~\ref{fig15:Maximov_n} represents power spectra for DF Tau
averaged from 2000 speckle images, recorded by the
intensifier-SensiCam (left) and PhotonMAX-512B (right) cameras.
The diffraction cutoff frequency of the aperture in the used
filter is marked by circles. The period of fringes in the power
spectrum corresponds to the angular separation between the
components of the binary (about 0.1$''$). Orientation of the
fringes determines the relative position angle of the binary
components; the magnitude difference between the two stars can be
found from the fringe contrast.

The drop of intensity with increasing spatial frequency on the
left image is caused by an additional photon noise spectrum,
resulting in a frequency-dependent bias in the fringe contrast.
Correcting this bias poses a separate intricate problem, being an
obstacle for differential photometry of the components of faint
sources. In the power spectra accumulated on the EMCCD data, the
influence of photon noise leads to a common change in the level of
background, which is easy to define. High contrast fringes are
indicative of the power spectrum of DF~Tau, obtained in 2007 with
the new system. These fringes can be traced up to the cutoff
frequency, determined by the diffraction at the telescope
aperture.

\section{CONCLUSIONS}

In conclusion, we shall list the main advantages of our new system
based on the EMCCD, as compared to the previous generation BTA
speckle interferometer based on an image intensifier optically
coupled to a CCD camera:

\begin{itemize}
\item[-] reduction of the total object exposure time thanks to the
4-fold performance advantage of the EMCCD; \item[-] improvement of
limiting magnitude up to 15$^m$ owing to the 10-fold quantum
efficiency advantage of EMCCD over the image intensifier
photocathode; \item[-] amelioration of image reconstruction
accuracy thanks to the absence of photon bias, which is the main
problem of image intensifiers application; \item[-] increase of
the maximal achievable magnitude difference in binary star
observations up to 5$^{m}$ owing to the high dynamic range, and
high signal-to-noise ratio in the reconstructed image; \item[-]
geometrical stability of the EMCCD-based detector and lack of
distortions inherent to image intensifiers; \item[-] broadening of
the speckle interferometer's working field up to 4$''$ owing to
the smaller optical magnification provided by the small EMCCD
pixel size (16 $\mu$m).
\end{itemize}

\begin{acknowledgments}

The authors are grateful to the professor of the Max Planck
Institute for Radio Astronomy in Bonn G.~Weigelt for his support
in procurance of some optical components of the speckle
interferometer. We would like to thank the head of the laboratory
of Advanced Design of the SAO RAS S.~V.~Markelov and the leading
engineer of the same laboratory V.A.~Murzin for useful discussions
and for their help in calibration measurements. We thank the
software engineers of the Information Technology Department
E.I.~Kaisina and S.L.~Komarinskii for their help in software
debugging. The work was supported by the program of the Presidium
of the RAS  ``The Origin, Structure and Evolution of the
Universe'', and the fundamental research program of the Division
of Physical Sciences of the RAS ``Extended Objects in the
Universe''. The work was as well partially supported by the grant
of the Russian Foundation of Basic Research (project no.
07-02-01489-a).

\end{acknowledgments}

\end{document}